\documentclass[10pt,a4paper]{article}
\usepackage{amsmath,amssymb,amsfonts}
\usepackage{graphicx}
\usepackage{hyperref}
\usepackage[margin=0.75in]{geometry}
\usepackage{float}
\usepackage{bm}

\usepackage{setspace}
\usepackage{titlesec}
\titlespacing*{\section}{0pt}{8pt}{4pt}
\titlespacing*{\subsection}{0pt}{6pt}{3pt}

\AtBeginDocument{
  \setlength{\abovedisplayskip}{4pt}
  \setlength{\belowdisplayskip}{4pt}
  \setlength{\abovedisplayshortskip}{2pt}
  \setlength{\belowdisplayshortskip}{2pt}
}

\let\oldbibliography\thebibliography
\renewcommand{\thebibliography}[1]{%
  \oldbibliography{#1}%
  \setlength{\itemsep}{0pt}%
  \setlength{\parskip}{0pt}%
}

\title{\texttt{dewi-kadita}: A Python Library for Idealized Fish Schooling Simulation with Entropy-Based Diagnostics}

\author{Sandy H. S. Herho$^{1,2,*}$, Iwan P. Anwar$^{3,4}$, Faruq Khadami$^{3}$, Alfita P. Handayani$^{5}$,\\ Karina A. Sujatmiko$^{3}$, Kamaluddin Kasim$^{6}$, Rusmawan Suwarman$^{7}$, and Dasapta E. Irawan$^{8}$}

\date{}

\begin{document}
\maketitle

\begin{center}
\small
$^{1}$Department of Earth and Planetary Sciences, University of California, Riverside, CA, USA\\
$^{2}$School of Systems Science and Industrial Engineering, State University of New York, Binghamton, NY, USA\\
$^{3}$Applied and Environmental Oceanography Research Group, Bandung Institute of Technology, Bandung, West Java, Indonesia\\
$^{4}$Samudera Sains Teknologi Ltd., Bandung, West Java, Indonesia\\
$^{5}$Spatial System and Cadaster Research Group, Bandung Institute of Technology, Bandung, West Java, Indonesia\\
$^{6}$Ministry of Maritime Affairs and Fisheries, North Jakarta, DKI Jakarta, Indonesia\\
$^{7}$Atmospheric Science Research Group, Bandung Institute of Technology, Bandung, West Java, Indonesia\\
$^{8}$Applied Geology Research Group, Bandung Institute of Technology, Bandung, West Java, Indonesia\\
$^{*}$e-mail: sandy.herho@email.ucr.edu
\end{center}

\begin{abstract}
\noindent Collective motion in fish schools exemplifies emergent self-organization in active matter systems, yet computational tools for simulating and analyzing these dynamics remain fragmented across research groups. We present \texttt{dewi-kadita}, an open-source Python library implementing the three-dimensional Couzin zone-based model with comprehensive entropy diagnostics tailored for marine collective behavior research. The library introduces seven information-theoretic metrics---school cohesion entropy, polarization entropy, depth stratification entropy, angular momentum entropy, nearest-neighbor entropy, velocity correlation entropy, and school shape entropy---that characterize distinct organizational features inaccessible to classical order parameters. These metrics combine into an Oceanic Schooling Index (OSI) providing a single scalar measure of collective disorder. Validation across four canonical configurations (swarm, torus, dynamic parallel, highly parallel) confirms correct reproduction of known phase behaviors: the swarm maintains disorder with polarization $P < 0.1$ and OSI $\approx 0.71$, while the highly parallel state achieves $P = 0.998$ with OSI $= 0.24$ and velocity correlation entropy vanishing to zero. The entropy framework successfully discriminates the torus and dynamic parallel configurations that exhibit comparable order parameter magnitudes through different organizational mechanisms. Numba just-in-time (JIT) compilation accelerates pairwise interaction calculations by $10$--$100\times$, enabling simulations of $150$--$250$ agents over $1000$--$2000$ time steps within five minutes on standard workstation hardware. NetCDF4 output ensures interoperability with oceanographic analysis tools. The library addresses the need for standardized, reproducible infrastructure in collective behavior modeling analogous to established molecular dynamics codes.
\end{abstract}


\section{Introduction}

Collective motion in biological systems represents a paradigmatic example of emergent complexity, wherein macroscopic spatiotemporal order arises spontaneously from local interactions among constituent agents \cite{Vicsek2012, Marchetti2013}. Schools of fish, flocks of birds, swarms of insects, and herds of mammals all exhibit coherent group-level behaviors that cannot be predicted from the properties of isolated individuals, placing these phenomena squarely within the domain of statistical physics and complex systems science \cite{Cavagna2014}. The theoretical understanding of such collective states has advanced substantially over the past three decades, driven by the recognition that active matter systems---composed of self-propelled particles that consume energy to generate motion---obey fundamentally different organizing principles than equilibrium thermodynamic systems \cite{Ramaswamy2010}.

Fish schooling constitutes perhaps the most extensively studied instance of biological collective motion, owing to the accessibility of laboratory experiments, the economic importance of commercial fisheries, and the striking visual regularity of natural schools \cite{Parrish1999, Partridge1982}. Empirical observations have established that schooling fish maintain characteristic nearest-neighbor distances, align their velocities with nearby conspecifics, and respond to perturbations through rapid information transfer across the group \cite{Katz2011, Herbert-Read2011}. These behavioral regularities motivated the development of agent-based models that distill the essential interaction rules governing collective dynamics while remaining analytically and computationally tractable \cite{Huth1992, Aoki1982}.

The seminal contribution of Vicsek \emph{et al.} \cite{Vicsek1995} introduced a minimal self-propelled particle model that captured the essential physics of flocking through velocity alignment interactions subject to angular noise. This model demonstrated that systems of active particles undergo a continuous phase transition from disordered to ordered collective motion as noise amplitude decreases, analogous to the ferromagnetic transition in spin systems. Subsequent theoretical analysis revealed that the Vicsek model belongs to a distinct universality class characterized by giant density fluctuations and long-range orientational correlations absent in equilibrium systems \cite{Toner1998, Chate2008}. The conceptual framework established by this work has informed virtually all subsequent modeling efforts in collective behavior.

Couzin \emph{et al.} \cite{Couzin2002} extended the Vicsek paradigm by introducing a zone-based behavioral architecture that more faithfully represents the graduated response of fish to neighbors at different distances. Their model partitions the space surrounding each agent into three concentric regions: a Zone of Repulsion (ZOR) governing collision avoidance, a Zone of Orientation (ZOO) mediating velocity alignment, and a Zone of Attraction (ZOA) generating long-range cohesion. This hierarchical structure, combined with a lexicographic priority rule that ensures collision avoidance supersedes social coordination, produces a rich phase diagram encompassing swarm, torus (milling), dynamic parallel, and highly parallel collective states \cite{Couzin2002, Couzin2005}. The Couzin model has since become a standard reference for theoretical and computational studies of fish schooling, with applications spanning behavioral ecology, robotics, and network science \cite{Sumpter2010, Brambilla2013}.

Despite the maturity of the theoretical framework, computational tools for simulating and analyzing Couzin model dynamics remain fragmented across research groups, with most implementations existing as unpublished laboratory codes lacking standardized interfaces, comprehensive diagnostics, or archival data formats \cite{Romanczuk2012}. This situation contrasts sharply with neighboring fields such as molecular dynamics, where community codes like LAMMPS and GROMACS provide validated, documented, and extensible platforms that accelerate scientific progress through reproducibility and interoperability \cite{Thompson2022, Abraham2015}. The absence of analogous infrastructure for collective behavior modeling impedes systematic parameter exploration, cross-study comparison, and integration with observational datasets from field studies \cite{Delcourt2013}.

A parallel limitation concerns the diagnostic metrics employed to characterize collective states. The classical order parameters---polarization $P$ quantifying velocity alignment and rotation $M$ measuring milling coherence---provide essential but incomplete descriptions of school organization \cite{Couzin2002, Tunstrom2013}. Two configurations with identical $(P, M)$ values may nonetheless differ substantially in their spatial structure, velocity correlations, or morphological characteristics. Information-theoretic measures based on Shannon entropy offer a complementary perspective by quantifying the spread or concentration of relevant distributions rather than their first moments \cite{Shannon1948, Crosato2018}. Entropy-based diagnostics have proven valuable in characterizing phase transitions, detecting early warning signals of critical slowing, and distinguishing superficially similar collective states in both simulations and experiments \cite{Attanasi2014, Cavagna2017}.

The application of entropy metrics to fish schooling has received limited attention despite the potential for extracting richer organizational signatures from simulation and observational data. Marine biologists have long recognized that schools exhibit complex internal structure---including density gradients, positional preferences, and shape variations---that influence predator evasion, foraging efficiency, and hydrodynamic performance \cite{Partridge1982, Hemelrijk2015}. Translating these biological insights into quantitative diagnostics requires metrics sensitive to the full distribution of relevant observables rather than summary statistics alone. Furthermore, the three-dimensional character of oceanic environments introduces vertical stratification and depth-dependent behaviors absent in two-dimensional laboratory studies, necessitating metrics that capture this additional dimensionality \cite{Handegard2012}.

Here we present \texttt{dewi-kadita}, an open-source Python library that provides a complete computational framework for three-dimensional Couzin model simulation with comprehensive entropy-based diagnostics tailored for marine collective behavior research. The library implements the full zone-based interaction architecture including visual field constraints, bounded turning rates, and stochastic perturbations, while introducing seven specialized entropy metrics that characterize distinct aspects of school organization: cohesion entropy based on nearest-neighbor distance distributions, polarization entropy quantifying velocity orientation spread, depth stratification entropy measuring vertical position uniformity, angular momentum entropy characterizing milling behavior, nearest-neighbor entropy capturing local density structure, velocity correlation entropy describing pairwise alignment distributions, and school shape entropy derived from principal component analysis of positional covariance. These metrics combine into an Oceanic Schooling Index (OSI) that provides a single scalar measure of collective disorder suitable for time series analysis and phase classification. The software architecture leverages Numba just-in-time (JIT) compilation \cite{Lam2015} for computational efficiency and produces Climate and Forecast (CF) compliant Network Common Data Format (NetCDF4) output \cite{Rew1990} ensuring interoperability with standard oceanographic analysis tools and adherence to Findable, Accessible, Interoperable, and Reusable (FAIR) data principles \cite{Wilkinson2016}.

Python's dominance in scientific computing offers mature numerical libraries and broad ecosystem  integration essential for research dissemination \cite{herho2025doublependulum, herho2025optionmc}. Numba JIT compilation eliminates performance concerns, achieving near-native speeds for pairwise interaction kernels while maintaining accessibility for rapid prototyping and collaborative development. This approach mirrors successful strategies in adjacent computational fields, positioning \texttt{dewi-kadita} as both production infrastructure and pedagogical tool for collective behavior research.

\section{Model Description}

The collective motion of fish schools constitutes a canonical example of self-organized behavior in biological systems, wherein macroscopic spatial order emerges spontaneously from local microscopic interactions among individuals \cite{Couzin2002, Vicsek2012}. This phenomenon belongs to the broader class of active matter systems, characterized by agents that consume energy to generate directed motion and interact through behavioral rules rather than conservative forces \cite{Marchetti2013}. The theoretical framework presented here derives the governing equations for zone-based collective dynamics following the seminal formulation of Couzin \emph{et al.} \cite{Couzin2002}, treating each fish as a self-propelled particle with constrained orientational degrees of freedom.

Consider an ensemble of $N$ self-propelled agents confined within a three-dimensional periodic domain $\Omega = [0, L]^3$ possessing toroidal topology. The imposition of periodic boundary conditions permits simulation of effectively unbounded oceanic environments while maintaining a finite computational domain, reflecting the observation that natural fish schools typically occupy spatial scales much smaller than the water bodies they inhabit \cite{Parrish1999}. The state of each agent $i \in \{1, 2, \ldots, N\}$ at discrete time $t$ is completely specified by the ordered pair $(\mathbf{r}_i(t), \mathbf{v}_i(t))$, where the position vector $\mathbf{r}_i \in \mathbb{R}^3$ resides on the three-torus $\mathbb{T}^3 \cong \mathbb{R}^3 / (L\mathbb{Z})^3$ and the velocity vector $\mathbf{v}_i \in \mathbb{R}^3$ determines both direction and magnitude of motion. The complete system state therefore inhabits the product manifold $\mathcal{M} = (\mathbb{T}^3)^N \times (\mathbb{R}^3)^N$.

A fundamental kinematic constraint governing fish locomotion is the maintenance of approximately constant swimming speed, reflecting physiological regulation mechanisms that decouple speed control from directional adjustment \cite{Videler1993}. We therefore impose the constant-speed constraint
\begin{equation}
\|\mathbf{v}_i(t)\| = v_0,
\label{eq:velocity_constraint}
\end{equation}
where $v_0 > 0$ denotes the constant cruising speed. Defining the unit velocity vector
\begin{equation}
\hat{\mathbf{v}}_i = \frac{\mathbf{v}_i}{\|\mathbf{v}_i\|},
\label{eq:unit_velocity}
\end{equation}
the constraint \eqref{eq:velocity_constraint} reduces the velocity degrees of freedom from $\mathbb{R}^3$ to the unit two-sphere $S^2 = \{\mathbf{x} \in \mathbb{R}^3 : \|\mathbf{x}\| = 1\}$, fundamentally altering the dynamical character of the system relative to passive Brownian particles and placing it within the class of self-propelled particle models \cite{Vicsek1995}.

The toroidal geometry necessitates careful treatment of interparticle distances. For agents $i$ and $j$ with positions $\mathbf{r}_i = (x_i, y_i, z_i)^\top$ and $\mathbf{r}_j = (x_j, y_j, z_j)^\top$, the displacement vector respecting periodic boundaries follows from the minimum image convention:
\begin{equation}
\boldsymbol{\Delta}_{ij} = \mathbf{r}_j - \mathbf{r}_i - L \cdot \mathrm{round}\left(\frac{\mathbf{r}_j - \mathbf{r}_i}{L}\right),
\label{eq:min_image}
\end{equation}
where the rounding operation $\mathrm{round}(\cdot)$ applies component-wise and maps each component to the nearest integer. The interparticle separation is then given by
\begin{equation}
d_{ij} = \|\boldsymbol{\Delta}_{ij}\| = \sqrt{\Delta_{ij,x}^2 + \Delta_{ij,y}^2 + \Delta_{ij,z}^2},
\label{eq:distance}
\end{equation}
which satisfies $d_{ij} \leq L\sqrt{3}/2$, corresponding to the maximum possible distance within the fundamental domain. This construction ensures that each agent perceives its neighbors through the topologically shortest path, a crucial requirement for correct simulation of periodic systems.

The behavioral response architecture introduced by Couzin \emph{et al.} \cite{Couzin2002} comprises three concentric spherical zones surrounding each focal individual, reflecting empirically observed gradations in fish response to neighbors at different distances \cite{Partridge1982, Huth1992}. The innermost region, designated the ZOR, is defined by
\begin{equation}
\mathrm{ZOR}_i = \{j \neq i : 0 < d_{ij} < r_r\},
\label{eq:zor_def}
\end{equation}
and governs collision avoidance behavior. When conspecifics intrude within the critical radius $r_r$, the focal agent experiences a repulsive stimulus that supersedes all other social interactions. The intermediate shell, termed the ZOO, spans
\begin{equation}
\mathrm{ZOO}_i = \{j \neq i : r_r \leq d_{ij} < r_o\},
\label{eq:zoo_def}
\end{equation}
and mediates velocity alignment interactions analogous to ferromagnetic spin coupling in condensed matter systems \cite{Vicsek1995}. The outermost region, the ZOA, encompasses
\begin{equation}
\mathrm{ZOA}_i = \{j \neq i : r_o \leq d_{ij} < r_a\},
\label{eq:zoa_def}
\end{equation}
and generates cohesive tendencies that maintain school integrity over larger spatial scales. The zone radii satisfy the strict hierarchy $0 < r_r < r_o \leq r_a$, ensuring unambiguous classification of all pairwise interactions.

Anatomical constraints on ocular positioning endow fish with a posterior blind region that must be incorporated into the interaction model \cite{Partridge1982}. We parametrize this limitation through a conical blind zone of half-angle $\alpha \in [0^{\circ}, 180^{\circ}]$ extending behind each agent along its negative velocity direction. Neighbor $j$ falls within the visual field of agent $i$ if and only if the angular criterion
\begin{equation}
\cos\theta_{ij} = \frac{\hat{\mathbf{v}}_i \cdot \boldsymbol{\Delta}_{ij}}{\|\boldsymbol{\Delta}_{ij}\|} > -\cos\alpha
\label{eq:visual_field}
\end{equation}
is satisfied, where $\theta_{ij}$ denotes the angle between the focal agent's heading and the direction toward neighbor $j$. The set of visible neighbors
\begin{equation}
\mathcal{V}_i = \left\{j \neq i : \frac{\hat{\mathbf{v}}_i \cdot \boldsymbol{\Delta}_{ij}}{\|\boldsymbol{\Delta}_{ij}\|} > -\cos\alpha\right\}
\label{eq:visible_set}
\end{equation}
thus excludes individuals located within the rear blind cone, introducing an essential asymmetry into the interaction topology. The limiting cases $\alpha = 0^{\circ}$ and $\alpha = 180^{\circ}$ correspond to complete omnidirectional vision and a hemispherical forward-only field, respectively.

The effective neighborhoods for each behavioral zone incorporate both distance and visibility constraints. The repulsion neighborhood is defined as
\begin{equation}
\mathcal{R}_i = \{j \neq i : d_{ij} < r_r \text{ and } j \in \mathcal{V}_i\},
\label{eq:repulsion_set}
\end{equation}
the orientation neighborhood as
\begin{equation}
\mathcal{O}_i = \{j \neq i : r_r \leq d_{ij} < r_o \text{ and } j \in \mathcal{V}_i\},
\label{eq:orientation_set}
\end{equation}
and the attraction neighborhood as
\begin{equation}
\mathcal{A}_i = \{j \neq i : r_o \leq d_{ij} < r_a \text{ and } j \in \mathcal{V}_i\}.
\label{eq:attraction_set}
\end{equation}

The response vectors for each behavioral zone derive from optimization principles governing fish behavior. Within the ZOR, agent $i$ seeks to maximize its distance from intruding neighbors. Defining the unit displacement vector
\begin{equation}
\hat{\boldsymbol{\Delta}}_{ij} = \frac{\boldsymbol{\Delta}_{ij}}{\|\boldsymbol{\Delta}_{ij}\|},
\label{eq:unit_displacement}
\end{equation}
the repulsion response vector is computed as
\begin{equation}
\mathbf{d}_i^{(r)} = -\sum_{j \in \mathcal{R}_i} \hat{\boldsymbol{\Delta}}_{ij}.
\label{eq:repulsion}
\end{equation}
The negative sign ensures motion directed away from neighbors, while the use of unit vectors prevents distant intruders from dominating the aggregate response.

For neighbors within the ZOO, the focal agent seeks to minimize angular deviation from the local consensus heading. This alignment tendency produces the orientation response vector
\begin{equation}
\mathbf{d}_i^{(o)} = \sum_{j \in \mathcal{O}_i} \hat{\mathbf{v}}_j,
\label{eq:orientation}
\end{equation}
where $\hat{\mathbf{v}}_j$ is the unit velocity of neighbor $j$ as defined in \eqref{eq:unit_velocity}. This summation of unit velocity vectors parallels the computation of magnetization in spin systems, with the resultant direction representing the locally averaged heading.

The attraction response governing behavior toward distant visible neighbors derives from cohesion optimization:
\begin{equation}
\mathbf{d}_i^{(a)} = \sum_{j \in \mathcal{A}_i} \hat{\boldsymbol{\Delta}}_{ij}.
\label{eq:attraction}
\end{equation}
This vector points toward the angular centroid of distant neighbors, generating the long-range cohesive force essential for maintaining school integrity.

The zone responses combine according to a lexicographic priority rule that reflects the biological primacy of collision avoidance \cite{Couzin2002, Krause2002}. The desired direction $\mathbf{d}_i$ governing agent $i$ takes the form
\begin{equation}
\mathbf{d}_i = \begin{cases}
\mathbf{d}_i^{(r)} & \text{if } \|\mathbf{d}_i^{(r)}\| > 0, \\[6pt]
w_o \mathbf{d}_i^{(o)} + \mathbf{d}_i^{(a)} & \text{if } \|\mathbf{d}_i^{(r)}\| = 0 \text{ and } (|\mathcal{O}_i| > 0 \text{ or } |\mathcal{A}_i| > 0), \\[6pt]
\hat{\mathbf{v}}_i & \text{otherwise},
\end{cases}
\label{eq:priority_rule}
\end{equation}
where $w_o \in [0,1]$ denotes the orientation weight parameter and $|\cdot|$ denotes set cardinality. This weighting factor modulates the relative influence of alignment versus cohesion forces when repulsion is absent, enabling continuous interpolation between milling-dominated dynamics at low $w_o$ and alignment-dominated behavior at high $w_o$. The priority structure ensures that collision avoidance always supersedes social coordination, consistent with the survival imperative. When no neighbors occupy any zone, the agent maintains its current heading.

The orientational dynamics of each agent are subject to biological constraints on angular velocity. Fish cannot execute instantaneous reorientation; rather, their turning rate is bounded by a maximum value determined by hydrodynamic and physiological factors \cite{Domenici2004}. Let $\theta_{\max} > 0$ denote the maximum angular displacement achievable per time step $\Delta t$. The constrained rotation from current direction $\hat{\mathbf{v}}_i$ toward the normalized desired direction
\begin{equation}
\hat{\mathbf{d}}_i = \frac{\mathbf{d}_i}{\|\mathbf{d}_i\|}
\label{eq:normalized_desired}
\end{equation}
proceeds as follows. First, compute the angular separation between current and target orientations:
\begin{equation}
\phi_i = \arccos\left(\mathrm{clip}(\hat{\mathbf{v}}_i \cdot \hat{\mathbf{d}}_i, -1, 1)\right),
\label{eq:angle_separation}
\end{equation}
where $\mathrm{clip}(x, a, b) = \max(a, \min(x, b))$ ensures numerical stability. The deterministic component of the updated direction then takes the form
\begin{equation}
\tilde{\mathbf{v}}_i = \begin{cases}
\hat{\mathbf{d}}_i & \text{if } \phi_i \leq \theta_{\max}, \\[6pt]
\displaystyle\frac{(1-\tau_i)\hat{\mathbf{v}}_i + \tau_i\hat{\mathbf{d}}_i}{\|(1-\tau_i)\hat{\mathbf{v}}_i + \tau_i\hat{\mathbf{d}}_i\|} & \text{otherwise},
\end{cases}
\label{eq:rotation_update}
\end{equation}
where the interpolation parameter
\begin{equation}
\tau_i = \frac{\theta_{\max}}{\phi_i}
\label{eq:interpolation_param}
\end{equation}
parametrizes partial rotation toward the target. This normalized linear interpolation (NLERP) scheme approximates spherical linear interpolation (SLERP) with error $\mathcal{O}(\theta_{\max}^3)$ for small turning angles, which is negligible for the turning rates typical of biological parameter regimes.

Environmental turbulence, sensory uncertainty, and intrinsic behavioral variability necessitate incorporation of stochastic perturbations into the orientational dynamics \cite{Couzin2002}. We model these fluctuations through additive Gaussian noise characterized by amplitude $\sigma$:
\begin{equation}
\hat{\mathbf{v}}_i^{\,\prime} = \frac{\tilde{\mathbf{v}}_i + \sigma\boldsymbol{\xi}_i}{\|\tilde{\mathbf{v}}_i + \sigma\boldsymbol{\xi}_i\|},
\label{eq:noise}
\end{equation}
where $\boldsymbol{\xi}_i = (\xi_{i,x}, \xi_{i,y}, \xi_{i,z})^\top$ denotes a three-dimensional random vector with independent standard normal components: $\xi_{i,k} \sim \mathcal{N}(0, 1)$ for $k \in \{x, y, z\}$. The normalization operation projects the perturbed vector back onto $S^2$, preserving the unit-speed constraint \eqref{eq:velocity_constraint}. In the small-noise limit $\sigma \ll 1$, this construction approximates isotropic diffusion on the unit sphere.

The positional dynamics follow simple Euler integration at constant speed along the instantaneous velocity direction:
\begin{equation}
\mathbf{r}_i(t + \Delta t) = \left[\mathbf{r}_i(t) + v_0 \hat{\mathbf{v}}_i^{\,\prime} \Delta t\right] \mod L,
\label{eq:position_update}
\end{equation}
where the modular arithmetic enforces periodic boundary conditions component-wise. The updated velocity is then
\begin{equation}
\mathbf{v}_i(t + \Delta t) = v_0 \hat{\mathbf{v}}_i^{\,\prime}.
\label{eq:velocity_update}
\end{equation}
The governing equations for the complete dynamical system thus comprise the coupled evolution rules \eqref{eq:rotation_update}--\eqref{eq:velocity_update} for orientation, velocity, and position of all $N$ agents.

The characterization of emergent collective states requires macroscopic observables that quantify global organizational structure \cite{Couzin2002, Vicsek2012}. The system possesses continuous rotational symmetry in velocity space: absent external fields or boundaries, no direction enjoys privileged status. Spontaneous breaking of this symmetry manifests as collective alignment and is captured by the polarization order parameter
\begin{equation}
P = \frac{1}{N}\left\|\sum_{i=1}^{N} \hat{\mathbf{v}}_i\right\|.
\label{eq:polarization}
\end{equation}
This quantity satisfies $P \in [0,1]$, with $P \to 0$ indicating an isotropic velocity distribution analogous to a paramagnetic phase and $P \to 1$ signifying complete alignment corresponding to ferromagnetic order. The polarization thus serves as the direct analog of magnetization in classical spin systems \cite{Vicsek1995}.

The system additionally exhibits rotational symmetry about the collective center of mass. Breaking of this symmetry corresponds to emergence of coherent angular momentum characteristic of milling or torus configurations. Let the group centroid be
\begin{equation}
\mathbf{r}_{\mathrm{cm}} = \frac{1}{N}\sum_{i=1}^{N} \mathbf{r}_i,
\label{eq:centroid}
\end{equation}
and define the unit vector from the centroid to agent $i$ as
\begin{equation}
\hat{\mathbf{r}}_i^{\,c} = \frac{\mathbf{r}_i - \mathbf{r}_{\mathrm{cm}}}{\|\mathbf{r}_i - \mathbf{r}_{\mathrm{cm}}\|}.
\label{eq:unit_centroid_vector}
\end{equation}
The rotation order parameter is then
\begin{equation}
M = \left\|\frac{1}{N}\sum_{i=1}^{N} \hat{\mathbf{r}}_i^{\,c} \times \hat{\mathbf{v}}_i\right\|,
\label{eq:rotation}
\end{equation}
which quantifies the coherence of rotational motion about the group center. When $M \to 1$, all agents circulate in phase about a common axis.

The two-dimensional order parameter space $(P, M)$ admits four qualitatively distinct collective phases identified by Couzin \emph{et al.} \cite{Couzin2002}. The \emph{swarm} phase, characterized by $P \approx 0$ and $M \approx 0$, exhibits disordered aggregation with neither translational nor rotational coherence, resembling a high-temperature paramagnetic state. The \emph{torus} phase, with moderate polarization but elevated rotation ($M \gtrsim 0.5$), displays coherent milling wherein agents execute circular orbits about a common center, breaking rotational but not translational symmetry. The \emph{dynamic parallel} phase, featuring high polarization ($P \gtrsim 0.7$) with residual fluctuations, represents aligned collective motion with significant directional variability analogous to a ferromagnet near criticality. The \emph{highly parallel} phase, distinguished by $P \gtrsim 0.9$ and negligible rotation, constitutes the ground state of the alignment-dominated regime with stable directed motion.

Beyond classical order parameters, information-theoretic measures provide complementary characterization of organizational complexity in collective systems \cite{Couzin2005}. For a discrete probability distribution $\{p_k\}_{k=1}^{n}$ obtained by histogramming a continuous observable into $n$ bins, the Shannon entropy
\begin{equation}
H = -\sum_{k=1}^{n} p_k \ln p_k
\label{eq:shannon}
\end{equation}
quantifies the uncertainty or disorder present in the distribution, where we adopt the convention $0 \ln 0 = 0$. The maximum entropy $H_{\max} = \ln n$ corresponds to a uniform distribution, and normalization by this maximum yields the dimensionless normalized entropy
\begin{equation}
\tilde{H} = \frac{H}{\ln n} \in [0, 1].
\label{eq:normalized_entropy}
\end{equation}
We now define seven specialized entropy metrics designed for characterizing marine collective behavior, each capturing distinct aspects of school organization.

The cohesion entropy characterizes the distribution of nearest-neighbor distances (NND), which reflects school compactness and predator defense efficiency \cite{Partridge1982}. For each agent $i$, the nearest-neighbor distance is
\begin{equation}
d_i^{\mathrm{NN}} = \min_{j \neq i} d_{ij},
\label{eq:nnd}
\end{equation}
computed using periodic boundary conditions via \eqref{eq:min_image}--\eqref{eq:distance}. The NND values $\{d_i^{\mathrm{NN}}\}_{i=1}^{N}$ are histogrammed into $n_c$ bins over the interval $[0, L/2]$, producing counts $\{c_k\}_{k=1}^{n_c}$ and probabilities $p_k = c_k / N$. The school cohesion entropy is then
\begin{equation}
H_{\mathrm{coh}} = \frac{-\sum_{k=1}^{n_c} p_k \ln p_k}{\ln n_c}.
\label{eq:cohesion_entropy}
\end{equation}
Tight schools exhibit narrow NND distributions yielding low $H_{\mathrm{coh}}$, while dispersed aggregations have broad distributions and high $H_{\mathrm{coh}}$.

The polarization entropy quantifies the spread of velocity orientations on the unit sphere, capturing alignment disorder distinct from the scalar polarization $P$ in \eqref{eq:polarization}. Converting unit velocities to spherical coordinates:
\begin{equation}
\theta_i = \arccos(\hat{v}_{i,z}), \quad \varphi_i = \arctan2(\hat{v}_{i,y}, \hat{v}_{i,x}) + \pi,
\label{eq:spherical_coords}
\end{equation}
where $\theta_i \in [0, \pi]$ is the polar angle and $\varphi_i \in [0, 2\pi]$ is the azimuthal angle. A two-dimensional histogram with $n_\theta$ polar bins and $n_\varphi$ azimuthal bins produces counts $h_{jk}$. To account for the non-uniform solid angle element $\sin\theta \, d\theta \, d\varphi$, the histogram is weighted by $\sin\theta_j$ where $\theta_j$ is the bin center:
\begin{equation}
\tilde{h}_{jk} = h_{jk} \sin\theta_j.
\label{eq:weighted_histogram}
\end{equation}
Normalizing to probabilities $p_{jk} = \tilde{h}_{jk} / \sum_{j,k} \tilde{h}_{jk}$ and flattening to a one-dimensional distribution, the polarization entropy is
\begin{equation}
H_{\mathrm{pol}} = \frac{-\sum_{j,k} p_{jk} \ln p_{jk}}{\ln(n_\theta n_\varphi)}.
\label{eq:polarization_entropy}
\end{equation}
Aligned schools have concentrated angular distributions (low $H_{\mathrm{pol}}$), while isotropic velocity fields approach maximum entropy.

Fish schools often stratify by depth following thermoclines, oxygen gradients, or light levels. The depth stratification entropy characterizes the uniformity of vertical position distribution. Histogramming the $z$-coordinates $\{z_i\}_{i=1}^{N}$ into $n_z$ bins over $[0, L]$ produces probabilities $\{p_k\}$, and the normalized entropy is
\begin{equation}
H_{\mathrm{depth}} = \frac{-\sum_{k=1}^{n_z} p_k \ln p_k}{\ln n_z}.
\label{eq:depth_entropy}
\end{equation}
A school concentrated at a single depth has low $H_{\mathrm{depth}}$; uniform vertical spread yields high $H_{\mathrm{depth}}$.

The angular momentum entropy characterizes the distribution of individual contributions to collective rotation, complementing the scalar rotation parameter $M$ in \eqref{eq:rotation}. For each agent, the magnitude of the cross product between its radial position (relative to centroid) and velocity direction quantifies its instantaneous contribution to rotational motion:
\begin{equation}
L_i = \|\hat{\mathbf{r}}_i^{\,c} \times \hat{\mathbf{v}}_i\| \in [0, 1].
\label{eq:individual_angular_momentum}
\end{equation}
Histogramming $\{L_i\}_{i=1}^{N}$ into $n_L$ bins over $[0, 1]$ and computing the normalized Shannon entropy yields
\begin{equation}
H_{\mathrm{ang}} = \frac{-\sum_{k=1}^{n_L} p_k \ln p_k}{\ln n_L}.
\label{eq:angular_entropy}
\end{equation}
Coherent milling produces uniform angular momentum contributions (high $H_{\mathrm{ang}}$ concentrated near $L_i \approx 1$), while mixed rotational states exhibit variable contributions.

Beyond first-nearest-neighbor distances, the variability of distances to multiple neighbors characterizes local density structure and social network topology. For each agent $i$, compute distances to its $k$ nearest neighbors (excluding self): $\{d_{i}^{(1)}, d_{i}^{(2)}, \ldots, d_{i}^{(k)}\}$. The coefficient of variation for agent $i$ is
\begin{equation}
\mathrm{CV}_i = \frac{\sigma_i}{\mu_i},
\label{eq:cv}
\end{equation}
where $\mu_i = k^{-1}\sum_{m=1}^{k} d_i^{(m)}$ and $\sigma_i = \sqrt{k^{-1}\sum_{m=1}^{k}(d_i^{(m)} - \mu_i)^2}$ are the mean and standard deviation of the $k$-NN distances. Histogramming $\{\mathrm{CV}_i\}_{i=1}^{N}$ into $n_{\mathrm{NN}}$ bins over $[0, 2]$ yields the entropy
\begin{equation}
H_{\mathrm{NN}} = \frac{-\sum_{k=1}^{n_{\mathrm{NN}}} p_k \ln p_k}{\ln n_{\mathrm{NN}}}.
\label{eq:nn_entropy}
\end{equation}
Regular spacing patterns produce low $\mathrm{CV}$ values with concentrated distributions (low $H_{\mathrm{NN}}$), while irregular local structures yield variable $\mathrm{CV}$.

The velocity correlation entropy characterizes the distribution of pairwise velocity alignments, capturing the full correlation structure beyond the mean captured by polarization. For all unique pairs $(i, j)$ with $i < j$, compute the velocity dot product:
\begin{equation}
\rho_{ij} = \hat{\mathbf{v}}_i \cdot \hat{\mathbf{v}}_j \in [-1, 1].
\label{eq:velocity_correlation}
\end{equation}
Shifting to the interval $[0, 1]$ via $\tilde{\rho}_{ij} = (\rho_{ij} + 1)/2$ and histogramming all $N(N-1)/2$ correlation values into $n_\rho$ bins yields the entropy
\begin{equation}
H_{\mathrm{vel}} = \frac{-\sum_{k=1}^{n_\rho} p_k \ln p_k}{\ln n_\rho}.
\label{eq:velocity_correlation_entropy}
\end{equation}
Highly aligned schools have all correlations near $\rho_{ij} \approx 1$ (low $H_{\mathrm{vel}}$), while disordered schools exhibit correlations spread across $[-1, 1]$ (high $H_{\mathrm{vel}}$).

Fish schools adopt various morphologies (spherical, elongated, disc-shaped) depending on behavioral state and environmental conditions \cite{Partridge1982}. The shape entropy characterizes this morphological variability through principal component analysis. Center the positions:
\begin{equation}
\tilde{\mathbf{r}}_i = \mathbf{r}_i - \mathbf{r}_{\mathrm{cm}},
\label{eq:centered_positions}
\end{equation}
and compute the $3 \times 3$ covariance matrix:
\begin{equation}
\mathbf{C} = \frac{1}{N}\sum_{i=1}^{N} \tilde{\mathbf{r}}_i \tilde{\mathbf{r}}_i^\top.
\label{eq:covariance}
\end{equation}
Let $\lambda_1 \geq \lambda_2 \geq \lambda_3 \geq 0$ be the eigenvalues of $\mathbf{C}$, representing variances along principal axes. Normalizing to a probability-like distribution:
\begin{equation}
q_m = \frac{\lambda_m}{\lambda_1 + \lambda_2 + \lambda_3}, \quad m \in \{1, 2, 3\},
\label{eq:eigenvalue_probs}
\end{equation}
the shape entropy is
\begin{equation}
H_{\mathrm{shape}} = \frac{-\sum_{m=1}^{3} q_m \ln q_m}{\ln 3}.
\label{eq:shape_entropy}
\end{equation}
Spherical schools have $\lambda_1 \approx \lambda_2 \approx \lambda_3$ yielding $H_{\mathrm{shape}} \to 1$, while highly elongated schools have $\lambda_1 \gg \lambda_2, \lambda_3$ yielding $H_{\mathrm{shape}} \to 0$.

The OSI provides a composite measure of school disorder by combining all seven entropy metrics with weights optimized for marine collective behavior assessment:
\begin{equation}
\mathrm{OSI} = \sum_{m=1}^{7} w_m H_m,
\label{eq:osi}
\end{equation}
where $H_m \in \{H_{\mathrm{coh}}, H_{\mathrm{pol}}, H_{\mathrm{depth}}, H_{\mathrm{ang}}, H_{\mathrm{NN}}, H_{\mathrm{vel}}, H_{\mathrm{shape}}\}$ and the weights satisfy $\sum_m w_m = 1$. The default weighting scheme emphasizes alignment and cohesion as primary schooling indicators:
\begin{equation}
(w_{\mathrm{coh}}, w_{\mathrm{pol}}, w_{\mathrm{depth}}, w_{\mathrm{ang}}, w_{\mathrm{NN}}, w_{\mathrm{vel}}, w_{\mathrm{shape}}) = (0.18, 0.28, 0.08, 0.08, 0.10, 0.18, 0.10).
\label{eq:osi_weights}
\end{equation}
The OSI satisfies $\mathrm{OSI} \in [0, 1]$, with $\mathrm{OSI} \to 0$ indicating highly ordered schooling (low entropy across all metrics) and $\mathrm{OSI} \to 1$ signifying disordered aggregation (high entropy).

A complementary scalar diagnostic is the Order Index (OI), which extracts the dominant ordering behavior from the standard order parameters:
\begin{equation}
\mathrm{OI} = \max(P, M).
\label{eq:order_index}
\end{equation}
This measure satisfies $\mathrm{OI} \in [0, 1]$, with high values indicating either strong alignment or coherent rotation.

\section{Numerical Implementation}

The theoretical framework established in the preceding section provides the mathematical foundation for computational simulation of fish schooling dynamics. We present here the numerical implementation embodied in the \texttt{dewi-kadita} library, a Python-based simulation package designed for high-performance computation of three-dimensional Couzin model dynamics with comprehensive diagnostic capabilities.

The computational architecture leverages several established scientific computing libraries within the Python ecosystem, each fulfilling specific functional roles. NumPy \cite{Harris2020} provides the fundamental data structures and numerical operations: multidimensional arrays store agent positions $\mathbf{r}_i$ and velocities $\mathbf{v}_i$, while vectorized functions compute Euclidean norms via \eqref{eq:distance}, dot products for angular separations in \eqref{eq:angle_separation} and visual field checks in \eqref{eq:visual_field}, cross products for the rotation order parameter in \eqref{eq:rotation}, and the modular arithmetic enforcing periodic boundary conditions in \eqref{eq:position_update}. Random number generation for initial conditions and stochastic perturbations in \eqref{eq:noise} employs NumPy's pseudorandom facilities.

SciPy \cite{Virtanen2020} supplies the $k$-d tree spatial data structure through its \texttt{cKDTree} class, which accelerates nearest-neighbor queries required for cohesion entropy computation via \eqref{eq:nnd} and $k$-NN entropy via \eqref{eq:cv}. The \texttt{boxsize} parameter enables proper handling of periodic boundary conditions, ensuring that distance calculations respect the minimum image convention in \eqref{eq:min_image}. This reduces algorithmic complexity from $\mathcal{O}(N^2)$ for exhaustive pairwise search to $\mathcal{O}(N \log N)$ for tree-based queries.

Matplotlib \cite{Hunter2007} generates all visualizations, including three-dimensional renderings of agent configurations using \texttt{Axes3D} projections with quiver plots for velocity vectors, time series of order parameters $P(t)$ and $M(t)$, and entropy diagnostic panels displaying all seven metrics. The library renders both static summary figures and individual frames for animated sequences.

Data serialization employs two complementary libraries. pandas \cite{McKinney2010} constructs \texttt{DataFrame} objects for tabular output, enabling structured export of order parameter time series, entropy metrics, and final agent states to comma-separated value (CSV) files. NetCDF4 \cite{Rew1990} writes self-describing binary files conforming to CF conventions, with proper dimension definitions (time, fish index, spatial coordinate), coordinate variables, and global attributes preserving complete simulation metadata for archival and interoperability with standard oceanographic analysis tools.

Pillow provides image processing capabilities for animation generation, applying Gaussian blur filters and brightness enhancement to create underwater visual effects in the rendered frames. tqdm displays progress bars during extended simulation runs and entropy computation loops, providing real-time feedback on computational progress.

The computational bottleneck in each time step is the $\mathcal{O}(N^2)$ evaluation of pairwise interactions required for zone membership determination and response vector accumulation in \eqref{eq:repulsion}--\eqref{eq:attraction}. Numba \cite{Lam2015} addresses this through JIT compilation of the core numerical kernels to optimized machine code. The velocity update function, which evaluates zone membership via \eqref{eq:zor_def}--\eqref{eq:zoa_def}, accumulates response vectors according to \eqref{eq:repulsion}--\eqref{eq:attraction}, applies the priority rule in \eqref{eq:priority_rule}, and performs constrained rotation via \eqref{eq:rotation_update}, is decorated with compilation directives specifying three operational modes. The \texttt{nopython=True} flag ensures complete compilation without fallback to the Python interpreter, eliminating interpreter overhead. The \texttt{cache=True} flag persists compiled kernels to disk, avoiding recompilation in subsequent sessions. The \texttt{parallel=True} flag combined with \texttt{prange} loop constructs enables automatic distribution of the outer loop over agents across available processor cores. This compilation strategy typically yields acceleration factors of 10--100$\times$ relative to pure Python implementations, rendering simulations of several hundred agents tractable on standard workstation hardware \cite{herho2025kh2d}.

The constrained rotation operation implements the spherical interpolation described by \eqref{eq:rotation_update} through normalized linear interpolation (NLERP). Given current direction $\hat{\mathbf{v}}_i$ and normalized desired direction $\hat{\mathbf{d}}_i$ from \eqref{eq:normalized_desired}, the algorithm first computes the angular separation $\phi_i$ according to \eqref{eq:angle_separation}. When $\phi_i \leq \theta_{\max}$, the agent achieves its desired orientation directly; otherwise, partial rotation proceeds with interpolation parameter $\tau_i$ from \eqref{eq:interpolation_param}. Stochastic perturbations enter through additive Gaussian fluctuations as specified in \eqref{eq:noise}, followed by renormalization to the unit sphere. Position updates follow first-order Euler integration of the advection dynamics given by \eqref{eq:position_update}.

The entropy-based diagnostic metrics require efficient computation of spatial statistics. Nearest-neighbor distance computation for \eqref{eq:nnd} employs the $k$-d tree with periodic boundary support. The velocity correlation entropy in \eqref{eq:velocity_correlation_entropy} requires evaluation of all $N(N-1)/2$ unique pairwise dot products, computed efficiently through matrix multiplication $\mathbf{V}\mathbf{V}^\top$ where $\mathbf{V}$ is the $N \times 3$ matrix of unit velocities, followed by extraction of upper-triangular elements. The polarization entropy in \eqref{eq:polarization_entropy} properly weights the two-dimensional angular histogram by $\sin\theta$ to account for the spherical coordinate Jacobian as specified in \eqref{eq:weighted_histogram}. The school shape entropy in \eqref{eq:shape_entropy} derives from eigenvalue decomposition of the position covariance matrix \eqref{eq:covariance}.

The \texttt{dewi-kadita} library includes four canonical test cases that systematically explore the phase space of collective behavior, selected to demonstrate the distinct dynamical regimes accessible to the Couzin model and to serve as validation benchmarks. These configurations correspond to well-characterized collective states identified in the original analysis \cite{Couzin2002}.

Case 1 (Swarm) employs parameters producing disordered aggregation: $N = 150$ agents, domain size $L = 25.0$, speed $v_0 = 1.0$, zone radii $(r_r, r_o, r_a) = (2.0, 3.0, 15.0)$, noise amplitude $\sigma = 0.2$ rad, turning rate $\theta_{\max} = 0.4$ rad, and blind angle $\alpha = 30^{\circ}$. The large repulsion zone relative to orientation prevents sustained alignment while maintaining loose cohesion, yielding low values of both order parameters ($P \approx 0$, $M \approx 0$). This configuration models feeding aggregations wherein collision avoidance dominates over coordination.

Case 2 (Torus) targets the milling regime through specialized initialization and parameter selection: $N = 200$ agents, $(r_r, r_o, r_a) = (1.0, 3.0, 10.0)$, reduced orientation weight $w_o = 0.1$ in \eqref{eq:priority_rule}, low noise $\sigma = 0.05$ rad, reduced turning rate $\theta_{\max} = 0.2$ rad, and large blind angle $\alpha = 150^{\circ}$. Agents begin arranged on a horizontal circle with tangential velocities to seed rotational motion. These settings suppress global alignment while preserving local attraction, yielding sustained rotation ($M \gtrsim 0.5$) with moderate polarization.

Case 3 (Dynamic Parallel) produces aligned collective motion with directional fluctuations: $N = 180$ agents, $L = 28.0$, $v_0 = 1.2$, $(r_r, r_o, r_a) = (1.0, 10.0, 18.0)$, $\sigma = 0.12$ rad, $\theta_{\max} = 0.35$ rad, $\alpha = 25^{\circ}$, and standard $w_o = 1.0$. The large orientation zone promotes alignment while moderate noise permits directional fluctuations, yielding high polarization ($P \gtrsim 0.7$) with quasi-periodic reorientation events characteristic of migrating pelagic schools.

Case 4 (Highly Parallel) achieves maximum alignment: $N = 250$ agents, $L = 35.0$, $v_0 = 1.5$, $(r_r, r_o, r_a) = (1.0, 12.0, 20.0)$, low noise $\sigma = 0.05$ rad, $\theta_{\max} = 0.2$ rad, and small blind angle $\alpha = 20^{\circ}$. These settings produce stable directed motion with polarization approaching unity ($P \gtrsim 0.9$) and minimal rotation, representing highly coordinated streaming behavior.

While these four test cases provide systematic coverage of the primary collective phases, the library supports arbitrary parameter configurations through text-based configuration files or direct Python API access. Users may specify custom values for all model parameters including zone radii $(r_r, r_o, r_a)$, kinematic constraints $(v_0, \theta_{\max})$, noise amplitude $\sigma$, blind angle $\alpha$, orientation weight $w_o$, ensemble size $N$, domain extent $L$, and temporal discretization $\Delta t$. Custom initialization protocols beyond standard random and toroidal configurations permit investigation of transient dynamics and relaxation behavior from prescribed initial conditions.

Output capabilities encompass multiple formats: time series of order parameters \eqref{eq:polarization}--\eqref{eq:rotation} and all entropy metrics \eqref{eq:cohesion_entropy}--\eqref{eq:shape_entropy} export to CSV files; complete trajectory data with full metadata write to CF-compliant NetCDF4 files; static visualizations render to PNG format; and animated three-dimensional trajectories generate as GIF sequences. The modular architecture separates core dynamics, diagnostic computation, data handling, and visualization into distinct subpackages, facilitating extension for specific research applications.

\section{Results and Discussion}

The \texttt{dewi-kadita} library was validated through systematic simulation of four canonical collective states, executed on a ThinkPad P52s workstation equipped with an Intel Core i7-8550U processor (4 cores, 8 threads at 4.0 GHz) running Fedora Linux 39. Each simulation completed within 5--6 minutes of wall-clock time, with the computational workload distributed as follows: main dynamics integration (9--25 s depending on ensemble size and duration), entropy metric computation (10--17 s), and visualization including animation generation (260--296 s). The Numba JIT compilation achieved the anticipated acceleration, rendering the $\mathcal{O}(N^2)$ pairwise interaction calculations tractable for ensembles of 150--250 agents over 1000--2000 time steps.

\begin{figure}[H]
\centering
\includegraphics[width=0.65\textwidth]{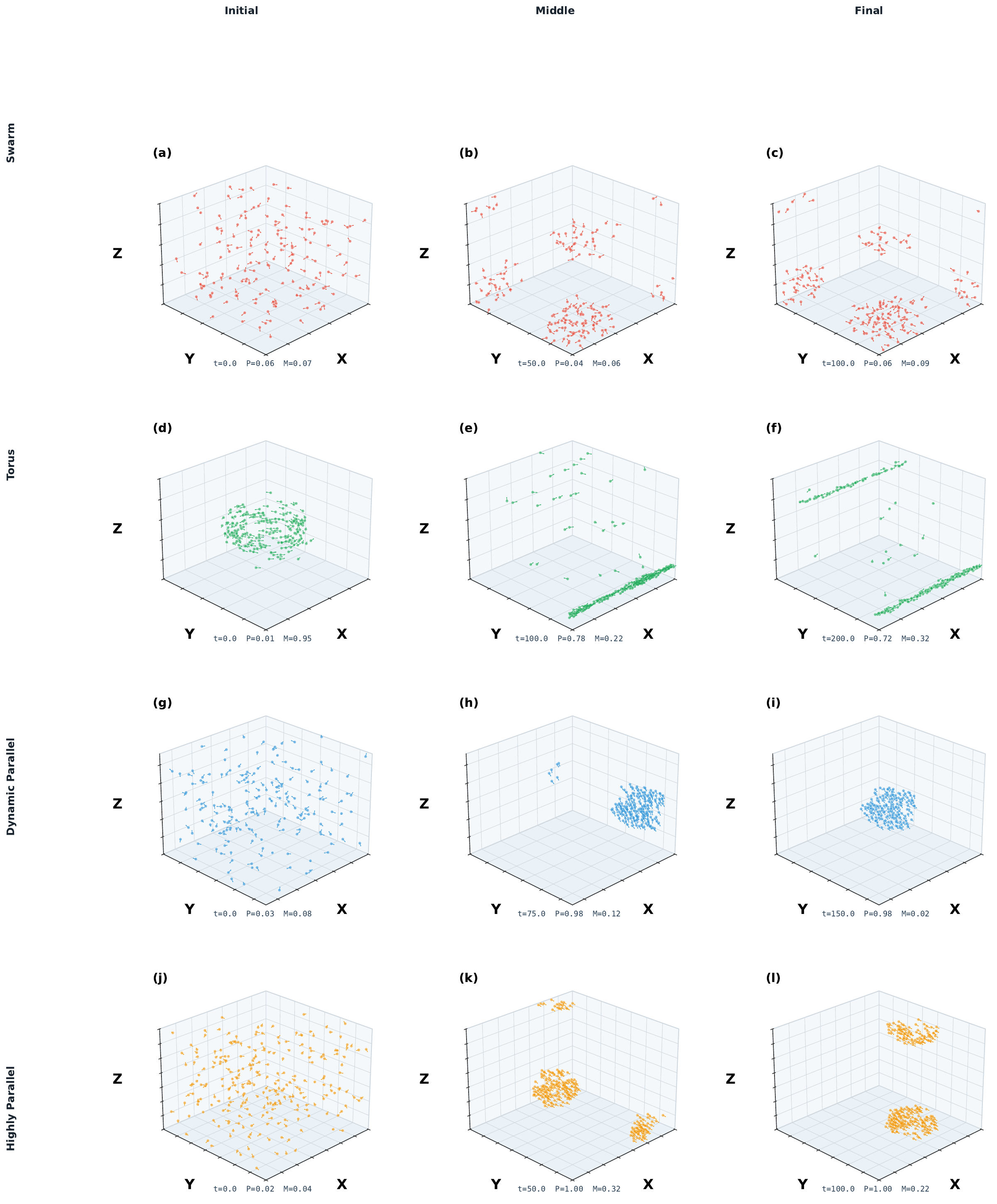}
\caption{Spatiotemporal evolution of collective states across four Couzin model scenarios. Rows correspond to distinct behavioral regimes: Swarm (a--c), Torus (d--f), Dynamic Parallel (g--i), and Highly Parallel (j--l). Columns represent temporal snapshots at initial ($t = 0$), middle ($t = T/2$), and final ($t = T$) simulation times. Three-dimensional quiver plots display fish positions with velocity vectors; instantaneous polarization $P$ and rotation $M$ values are annotated for each panel. The Swarm configuration maintains disordered aggregation throughout, while the Torus case transitions from initial milling ($M = 0.95$) toward partial alignment. Both Dynamic Parallel and Highly Parallel cases exhibit rapid ordering from random initial conditions to coherent directed motion ($P > 0.98$).}
\label{fig:spatiotemporal}
\end{figure}

Figure~\ref{fig:spatiotemporal} presents the spatiotemporal evolution of agent configurations across all four scenarios. The Swarm case (panels a--c) exhibits persistent disorder characteristic of the paramagnetic-like phase, with both order parameters remaining below 0.1 throughout the $T = 100$ time unit simulation. This behavior emerges from the combination of a relatively large repulsion zone ($r_r = 2.0$) that dominates the narrow orientation zone ($r_o - r_r = 1.0$), elevated noise amplitude ($\sigma = 0.20$ rad), and high turning rate ($\theta_{\max} = 0.40$ rad). The resulting dynamics prevent sustained alignment while maintaining loose cohesion through the attraction zone, consistent with theoretical predictions for this parameter regime \cite{Tunstrom2013}. Spatial statistics reveal that nearest-neighbor distances stabilize around $1.73 \pm 0.31$ body lengths by the final state, indicating effective collision avoidance without crystalline ordering.

The Torus scenario (panels d--f) was initialized with agents arranged on a horizontal circle possessing tangential velocities, yielding an initial rotation parameter $M = 0.95$. However, the system did not maintain stable milling; instead, it evolved toward a mixed state with final values $P = 0.72$ and $M = 0.32$. This transition reflects the inherent instability of the torus configuration under the chosen parameters: although the large blind angle ($\alpha = 150^{\circ}$) and reduced orientation weight ($w_o = 0.1$) favor following behavior, the finite system size and periodic boundary conditions permit spontaneous symmetry breaking toward translational order \cite{Calovi2014}. The polarization entropy decreased from $H_{\mathrm{pol}} = 0.67$ initially to $H_{\mathrm{pol}} = 0.39$ at equilibrium, quantifying the partial alignment that emerged despite parameter settings intended to suppress it. This result underscores that the Couzin model phase boundaries are not sharp and that finite-size effects can drive transitions between nominally distinct collective states \cite{Barberis2016}.

\begin{figure}[H]
\centering
\includegraphics[width=0.85\textwidth]{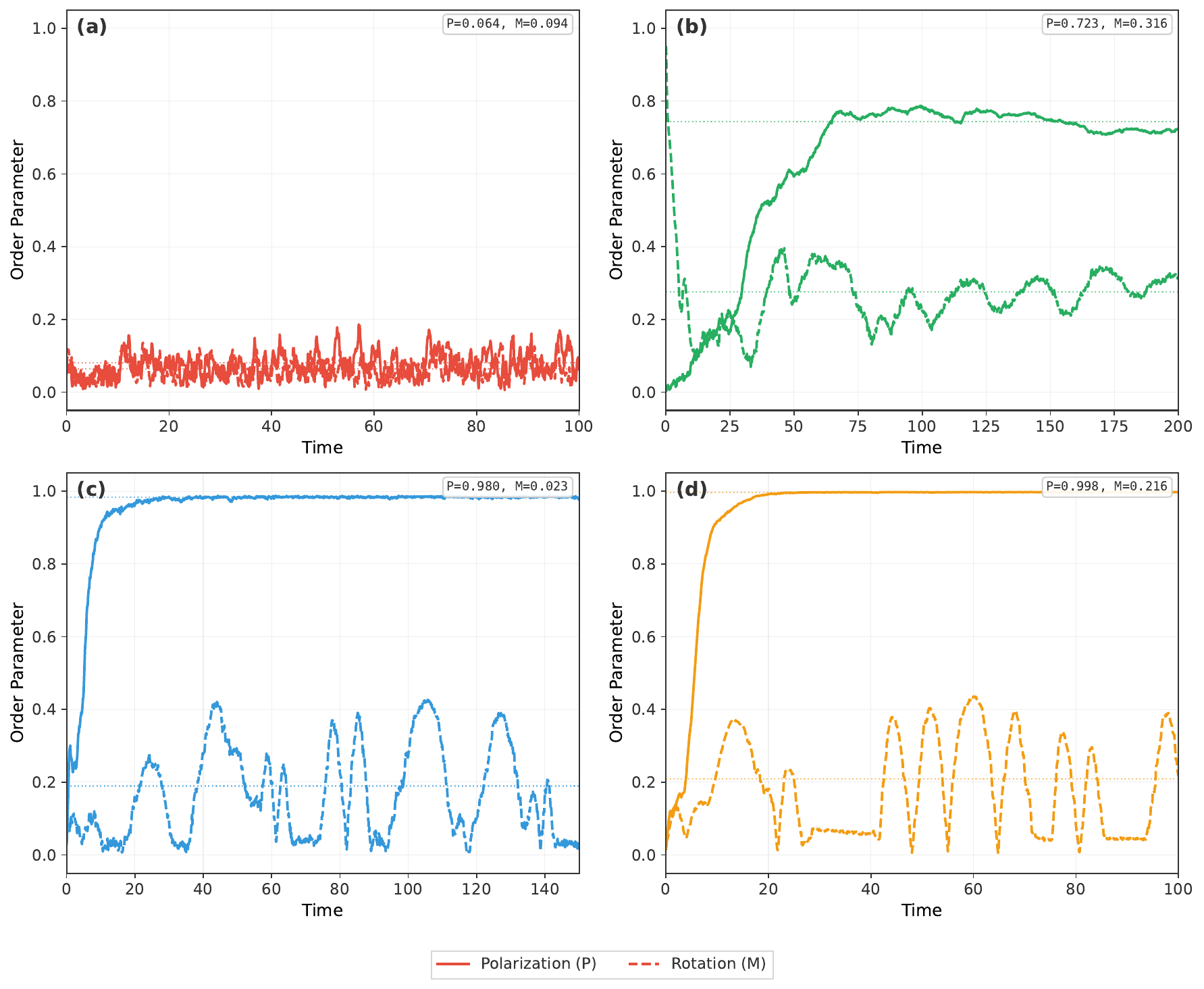}
\caption{Temporal evolution of classical order parameters for all four scenarios. Each panel displays polarization $P$ (solid lines) and rotation $M$ (dashed lines) as functions of time, with horizontal dotted lines indicating equilibrium mean values computed over the second half of each simulation. Panel annotations report final values. (a) Swarm: both parameters fluctuate near zero with no sustained ordering. (b) Torus: initial high rotation decays while polarization grows, indicating transition from milling toward partial alignment. (c) Dynamic Parallel: rapid convergence to $P \approx 0.98$ within $t \approx 30$ time units. (d) Highly Parallel: fastest convergence to near-unity polarization ($P = 0.998$) with minimal fluctuations ($\sigma_P = 0.0003$).}
\label{fig:order_parameters}
\end{figure}

The Dynamic Parallel and Highly Parallel configurations (Figure~\ref{fig:spatiotemporal}, panels g--l) both achieved strong alignment from random initial conditions, but with distinct dynamical signatures visible in Figure~\ref{fig:order_parameters}. The Dynamic Parallel case converged to $P = 0.98$ with residual fluctuations ($\sigma_P = 0.0018$), while the Highly Parallel case attained $P = 0.998$ with fluctuations suppressed by an order of magnitude ($\sigma_P = 0.0003$). This difference arises primarily from the noise amplitude: $\sigma = 0.12$ rad versus $\sigma = 0.05$ rad. The larger orientation zones in both cases ($r_o = 10$ and $r_o = 12$, respectively) ensure that each agent averages over many neighbors when computing its desired heading, providing the long-range velocity correlations necessary for global alignment \cite{Ginelli2015}. The school morphology in the Highly Parallel case (panel l) shows pronounced elongation along the direction of motion, with school shape entropy $H_{\mathrm{shape}} = 0.28$ indicating strong anisotropy, compared to $H_{\mathrm{shape}} = 0.99$ for the nearly spherical Dynamic Parallel school.

\begin{figure}[H]
\centering
\includegraphics[width=0.85\textwidth]{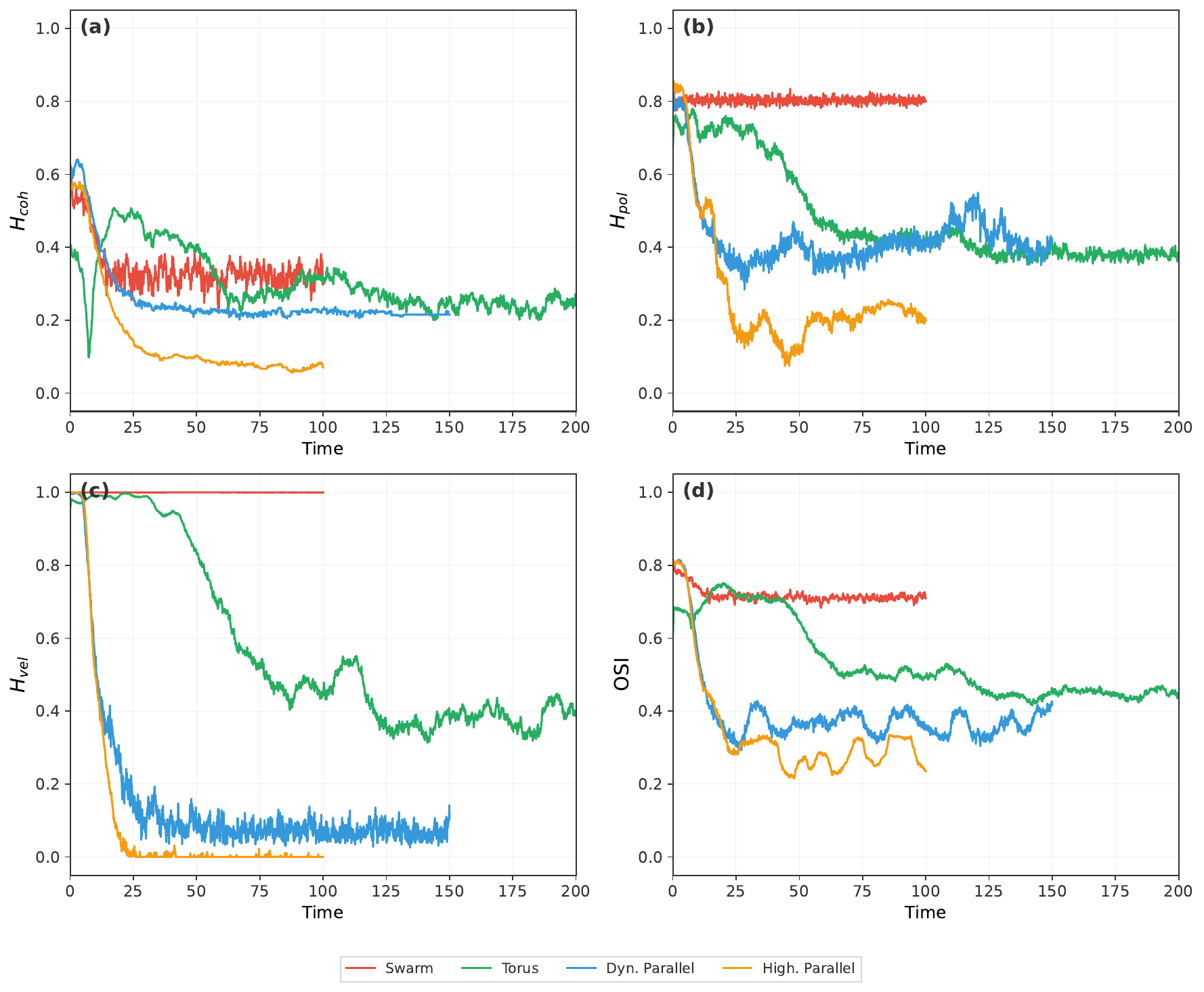}
\caption{Temporal evolution of oceanic entropy metrics across all scenarios. (a) School cohesion entropy $H_{\mathrm{coh}}$ based on nearest-neighbor distance distributions. (b) Polarization entropy $H_{\mathrm{pol}}$ quantifying velocity orientation spread on the unit sphere. (c) Velocity correlation entropy $H_{\mathrm{vel}}$ characterizing pairwise alignment distributions. (d) OSI, a weighted composite of all seven entropy metrics. Lower values indicate more ordered states. The Highly Parallel case achieves the lowest final OSI (0.24), while Swarm maintains elevated disorder (OSI $\approx 0.71$) throughout.}
\label{fig:entropy_evolution}
\end{figure}

The entropy-based diagnostics implemented in \texttt{dewi-kadita} provide complementary characterization of collective organization beyond the classical order parameters. Figure~\ref{fig:entropy_evolution} displays the temporal evolution of four key metrics. The velocity correlation entropy $H_{\mathrm{vel}}$ proves particularly discriminating: it drops from near-unity (indicating uniform distribution of pairwise correlations across $[-1, 1]$) to effectively zero in the Highly Parallel case, signifying that all velocity pairs have become perfectly correlated ($\rho_{ij} \approx 1$). This metric captures information absent from the scalar polarization $P$; two configurations with identical $P$ values could exhibit different $H_{\mathrm{vel}}$ if their correlation distributions differ in shape \cite{Attanasi2014}. The Swarm case maintains $H_{\mathrm{vel}} \approx 1.0$ throughout, confirming the absence of velocity correlations beyond random expectation.

The cohesion entropy $H_{\mathrm{coh}}$ reveals distinct spacing patterns across scenarios. The Swarm equilibrates at $H_{\mathrm{coh}} = 0.31$, reflecting a relatively narrow nearest-neighbor distance distribution centered around the repulsion radius. The Highly Parallel case achieves $H_{\mathrm{coh}} = 0.07$, indicating highly uniform spacing---a consequence of the balance between collision avoidance and the compressive effect of cohesive attraction acting perpendicular to the direction of collective motion \cite{Katz2011}. Interestingly, the Torus case exhibits intermediate cohesion entropy ($H_{\mathrm{coh}} = 0.25$) with elevated nearest-neighbor variability (coefficient of variation entropy $H_{\mathrm{NN}} = 0.60$), reflecting the density gradients inherent in rotating configurations where agents on the outer edge experience different local environments than those near the center.

\begin{figure}[H]
\centering
\includegraphics[width=0.85\textwidth]{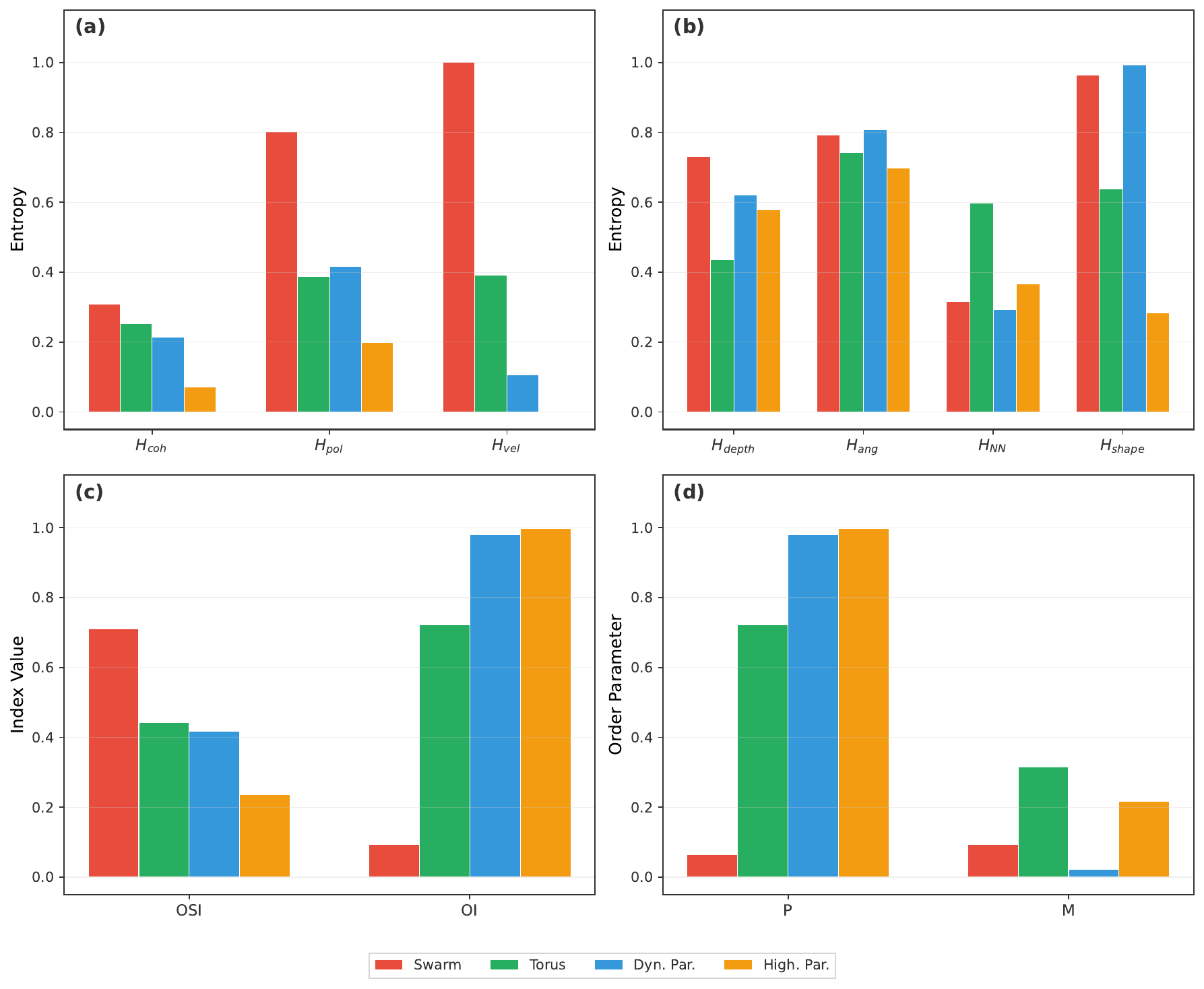}
\caption{Final state comparison of all metrics across scenarios. (a) Primary entropy metrics: school cohesion ($H_{\mathrm{coh}}$), polarization ($H_{\mathrm{pol}}$), and velocity correlation ($H_{\mathrm{vel}}$). (b) Secondary entropy metrics: depth stratification ($H_{\mathrm{depth}}$), angular momentum ($H_{\mathrm{ang}}$), nearest-neighbor variability ($H_{\mathrm{NN}}$), and school shape ($H_{\mathrm{shape}}$). (c) Composite indices: OSI and OI. (d) Classical order parameters: polarization ($P$) and rotation ($M$). The entropy metrics collectively distinguish all four phases, with OSI providing monotonic ranking from ordered (Highly Parallel, OSI $= 0.24$) to disordered (Swarm, OSI $= 0.71$).}
\label{fig:entropy_profile}
\end{figure}

Figure~\ref{fig:entropy_profile} summarizes the equilibrium entropy profiles, demonstrating that the seven-metric framework successfully discriminates all four collective phases. The OSI ranks the scenarios in intuitive order: Highly Parallel (OSI $= 0.24$) $<$ Dynamic Parallel (OSI $= 0.42$) $<$ Torus (OSI $= 0.44$) $<$ Swarm (OSI $= 0.71$). This ranking aligns with but refines the OI classification, which cannot distinguish the Torus from Dynamic Parallel cases since both achieve comparable maximum order parameters through different mechanisms (rotation versus polarization). The depth stratification entropy $H_{\mathrm{depth}}$ remains moderate (0.43--0.73) across all scenarios, reflecting the absence of vertical gradients in our homogeneous periodic domain; this metric would become informative in simulations incorporating bathymetry, thermoclines, or predator avoidance \cite{Handegard2012}.

The angular momentum entropy $H_{\mathrm{ang}}$ provides insight into the heterogeneity of rotational contributions within each school. The Torus case exhibits $H_{\mathrm{ang}} = 0.74$, lower than the Swarm value of 0.79, indicating more uniform individual contributions to collective rotation despite the decay of global milling. This suggests that local rotational correlations persist even as the macroscopic torus structure dissolves. The Dynamic Parallel case shows elevated $H_{\mathrm{ang}} = 0.81$ at equilibrium with high temporal variability ($\sigma = 0.17$), consistent with sporadic directional changes that transiently generate rotational motion before the school realigns \cite{Herbert-Read2011}.

The default weighting scheme for the OSI (Equation~\ref{eq:osi_weights}) assigns highest priority to polarization entropy ($w_{\mathrm{pol}} = 0.28$), reflecting the biological primacy of alignment in schooling fish \cite{Lopez2012}. However, the modular architecture of \texttt{dewi-kadita} permits user-specified weights for applications where other organizational features dominate. For instance, studies of vertical migration might elevate $w_{\mathrm{depth}}$, while investigations of milling behavior could increase $w_{\mathrm{ang}}$. The library outputs all seven component entropies separately, enabling post-hoc reweighting without rerunning simulations.

The computational experiments reveal several noteworthy features of the Couzin model dynamics under periodic boundary conditions. First, the convergence timescales vary substantially across phases: the Highly Parallel case achieves 99\% of its equilibrium polarization within $t \approx 20$ time units, while the Torus case requires the full $T = 200$ simulation to approach quasi-equilibrium due to the competing tendencies of milling and alignment. Second, the periodic boundaries prevent true school fragmentation, which has been observed experimentally in open domains when noise or perturbations exceed critical thresholds \cite{Tunstrom2013}. Third, the stochastic dynamics introduce measurable fluctuations even at equilibrium: the Swarm case exhibits coefficient of variation $\mathrm{CV}_P = 0.45$ for polarization, while the Highly Parallel case achieves $\mathrm{CV}_P = 0.0003$, spanning three orders of magnitude in relative stability.

The CF-compliant NetCDF4 output format ensures interoperability with standard oceanographic analysis tools, facilitating integration of \texttt{dewi-kadita} simulations with observational datasets. Each output file contains complete trajectory data (positions and velocities at all saved time steps), time series of order parameters and entropy metrics, and comprehensive metadata including all model parameters, random seeds, and software versions. This archival completeness supports reproducibility standards increasingly required by funding agencies and journals \cite{Stodden2016}. The CSV output option provides immediate accessibility for users unfamiliar with NetCDF4 conventions, while the animated GIF sequences enable rapid visual assessment of simulation quality.

The present validation confirms that \texttt{dewi-kadita} correctly reproduces the four canonical collective phases identified in the original Couzin formulation \cite{Couzin2002}, while extending the diagnostic framework through information-theoretic metrics tailored for marine applications. The library's Numba-accelerated core achieves sufficient performance for parameter sweeps and sensitivity analyses on standard workstation hardware, though users investigating larger ensembles ($N > 500$) or extended durations may benefit from the parallel loop constructs enabled by the \texttt{parallel=True} compilation flag. Future development priorities include incorporation of heterogeneous agent populations \cite{Jolles2017}, external flow fields representing oceanic currents, and predator--prey interaction modules for studying collective escape responses.

\section{Conclusion}

The \texttt{dewi-kadita} library provides a validated, open-source platform for three-dimensional Couzin model simulation that extends classical order parameter diagnostics through a comprehensive entropy-based framework tailored for marine collective behavior research. Systematic validation across four canonical collective states---swarm, torus, dynamic parallel, and highly parallel---confirms correct reproduction of known phase behaviors while demonstrating that the seven oceanic entropy metrics successfully discriminate configurations indistinguishable by polarization and rotation alone. The OSI furnishes a single scalar measure that ranks collective disorder monotonically across phases, enabling automated state classification and transition detection in long time series. Numba-accelerated computation renders parameter sweeps tractable on standard workstation hardware, while CF-compliant NetCDF4 output ensures interoperability with oceanographic analysis pipelines and adherence to contemporary data sharing standards. The modular architecture accommodates straightforward extension toward heterogeneous agent populations, external flow fields, and predator--prey dynamics, positioning the library as foundational infrastructure for computational studies bridging statistical physics and marine ecology.

\section*{Acknowledgments}

S.H.S.H. acknowledges financial support from the Dean's Distinguished Fellowship 2023 at the University of California, Riverside. F.K. and I.P.A. acknowledge support from Bandung Institute of Technology through the Research, Community Service and Innovation Program (Grant No. FITB.PPMI-1-04-2025) and the Early Career Research Scheme administered by the Directorate of Research and Innovation (Grant No. 2352/IT1.B07.1/TA.00/2025).

\section*{Author Contributions}

S.H.S.H.: Conceptualization; Formal analysis; Investigation; Methodology; Software; Validation; Visualization; Funding acquisition; Writing -- original draft; Project administration. I.P.A.: Methodology; Resources; Funding acquisition; Supervision; Writing -- review \& editing. F.K.: Funding acquisition; Supervision; Writing -- review \& editing. A.P.H.: Funding acquisition; Supervision; Writing -- review \& editing. K.A.S.: Funding acquisition; Supervision; Writing -- review \& editing. K.K.: Resources; Writing -- review \& editing. R.S.: Funding acquisition; Supervision; Writing -- review \& editing. D.E.I.: Funding acquisition; Supervision; Writing -- review \& editing. All authors have read and agreed to the published version of the manuscript.

\section*{Open Research}

The \texttt{dewi-kadita} library source code is hosted on GitHub at \url{https://github.com/sandyherho/dewi-kadita} under the MIT license and is available for installation via PyPI at \url{https://pypi.org/project/dewi-kadita/}. Supplementary materials including Python scripts for data analysis and figure generation are provided at \url{https://github.com/sandyherho/suppl-dewi-kadita}. All simulation outputs---comprising NetCDF4 trajectory files, order parameter time series, entropy metric data, static figures, animated visualizations, computational logs, and statistical reports---are permanently archived on the Open Science Framework at \url{https://doi.org/10.17605/OSF.IO/GP2XU} and released under the MIT license.

\end{document}